\documentclass[sigconf,authorversion]{acmart}

\AtBeginDocument{%
 }

\usepackage[most]{tcolorbox}
\usepackage{xcolor}

\definecolor{themeRed}{HTML}{F25050}
\definecolor{themeBlue}{HTML}{506AF2}

\tcbset{
  highlights/.style={
    enhanced,
    colframe=blue!30,
    colback=blue!60,
    boxrule=0pt,
    corners=south,
    rounded corners=north,
    arc=1mm,
    boxsep=0pt,
    left=1mm,
    right=1mm,
    top=0.5mm,
    bottom=0.5mm, fontupper=\bfseries\color{white},
  },
}
\newtcolorbox{custombox}[1]{
	colback=gray!10,
	colframe=gray!70,
	left=1mm,
	right=1mm,
	top=1mm,
	bottom=1mm,
	fonttitle=\bfseries,
	arc=0mm,
	leftrule=1mm,
	rightrule=0mm,
	toprule=0mm,
	bottomrule=0mm,
	notitle,
	before=\par\smallskip\noindent,
	before upper={\textbf{#1: } },
}

\copyrightyear{2025}
\acmYear{2025}
\setcopyright{rightsretained}
\acmConference[SIGCSE TS 2025]{Proceedings of the 56th ACM Technical
Symposium on Computer Science Education V. 2}{February 26-March 1,
2025}{Pittsburgh, PA, USA}
\acmBooktitle{Proceedings of the 56th ACM Technical Symposium on Computer
Science Education V. 2 (SIGCSE TS 2025), February 26-March 1, 2025,
Pittsburgh, PA, USA}
\acmDOI{10.1145/3641555.3705266}
\acmISBN{979-8-4007-0532-8/25/02}

\begin{document}


\title{The Evolving Usage of GenAI by Computing Students}


\author{Irene Hou}
\affiliation{
  \institution{Temple University}
  \city{Philadelphia}
  \state{PA}
  \country{United States}}
\email{irene.hou@temple.edu}
\orcid{0009-0008-0511-7685}

\author{Hannah Vy Nguyen}
\affiliation{
  \institution{Temple University}
  \city{Philadelphia}
  \state{PA}
  \country{United States}}
\email{hannah.nguyen0002@temple.edu}
\orcid{0009-0003-7422-1797}

\author{Owen Man}
\affiliation{
  \institution{Temple University}
  \city{Philadelphia}
  \state{PA}
  \country{United States}}
\email{owen.man@temple.edu}
\orcid{0009-0003-0527-1395}

\author{Stephen MacNeil}
\affiliation{%
  \institution{Temple University}
  \city{Philadelphia}
  \state{PA}
  \country{United States}}
\email{stephen.macneil@temple.edu}
\orcid{0000-0003-2781-6619}

\renewcommand{\shortauthors}{Irene Hou, Hannah Vy Nguyen, Owen Man, and Stephen MacNeil}

\begin{abstract}

Help-seeking is a critical aspect of learning and problem-solving for computing students. Recent research has shown that many students are aware of generative AI (GenAI) tools; however, there are gaps in the extent and effectiveness of how students use them. With over two years of widespread GenAI usage, it is crucial to understand whether students' help-seeking behaviors with these tools have evolved and how. This paper presents findings from a repeated cross-sectional survey conducted among computing students across North American universities (\textit{n=95}). Our results indicate shifts in GenAI usage patterns. In 2023, 34.1\% of students (\textit{n=47}) reported never using ChatGPT for help, ranking it fourth after online searches, peer support, and class forums. By 2024, this figure dropped sharply to 6.3\% (\textit{n=48}), with ChatGPT nearly matching online search as the most commonly used help resource. Despite this growing prevalence, there has been a decline in students’ hourly and daily usage of GenAI tools, which may be attributed to a common tendency to underestimate usage frequency. These findings offer new insights into the evolving role of GenAI in computing education, highlighting its increasing acceptance and solidifying its position as a key help resource.

\end{abstract}



\keywords{Generative AI, ChatGPT, help-seeking, computing education}



\maketitle

\section{Context and Motivation}

Since the introduction of mainstream generative AI (GenAI) tools, such as ChatGPT and Github Copilot, researchers have been surveying students to understand the extent to which students are using them~\cite{hou2024effects, prather2023robots, padiyath2024insights, smith2024early}. However, there lacks a longitudinal analysis of student usage to understand how it might be changing as students become more familiar with GenAI tools. Capturing this changing nature is crucial, as early studies revealed a clear gap between early and late adopters of GenAI tools~\cite{hou2024effects, smith2024early}. These gaps are important to evaluate because they relate to the recently observed divide in student performance~\cite{prather2024widening}. Consequently, we urgently need updates to these surveys with results that investigate longitudinal trends to ensure that the usage gap is closing rather than widening. 

\section{Methodology and Results}

To gain insight into evolving student use of GenAI tools over time, we conducted a repeated cross-sectional survey. Participants for the first survey were recruited between July-October 2023 from North American universities, totaling 47 computing students (13 women, 32 men, 2 preferred not to say). Participants for the second survey were recruited between June-September 2024 from North American universities, totaling 48 computing students (17 women, 29 men, 2 non-binary), including 2 students from European universities. All participants were recruited using the same advertisement by computing student organization leaders and faculty. Participants from the two groups were distributed across first through fourth-year students, totaling 2 incoming first-years, 31 first-years, 25 second-years, 23 third-years, 10 fourth-years, and 3 recent graduates. As compensation, participants were entered in a drawing for a \$25 gift card. The research was conducted following approval from our university's Institutional Review Board.

\begin{table}[h!]
\centering
\caption{Percentage Point Differences in Frequency of Resource Usage by Respondents between 2023-2024 (\textit{n=47, n=48})}
\scalebox{0.68}{
\begin{tabular}{l c c c c c}
\toprule
\textbf{Resource}        & \textbf{Hourly (\%)} & \textbf{Daily (\%)}   & \textbf{Weekly (\%)}  & \textbf{Monthly (\%)}   & \textbf{Never (\%)}    \\ \midrule
Online search   & -0.40   & -9.40  & +12.01   & -4.26 &  +2.04  \\
GenAI (e.g. ChatGPT)        & -8.51   & +8.02  & +37.10   & -8.82 & -27.79 \\
Friends         & -4.26   & -8.91  & +20.12   & -8.78 &  +1.82  \\
Class forum     & -4.26   & -2.53  & +13.96   & -4.52 & -2.66  \\
Instructor      & -2.13   & -6.52  & +7.71    & -9.09 & +10.02 \\
TA              & -2.13   & -4.30  & +13.92   & -11.13 & +3.63 \\
GitHub Copilot  & -0.04   &  +2.04  & -0.13    &  +4.08  & -5.94  \\ \bottomrule
\end{tabular}
}
\label{tab:frequent-percent}
\end{table}

\begin{figure}
    \setlength{\belowcaptionskip}{-10pt} 
    \centering
    \includegraphics[width=0.9\linewidth]{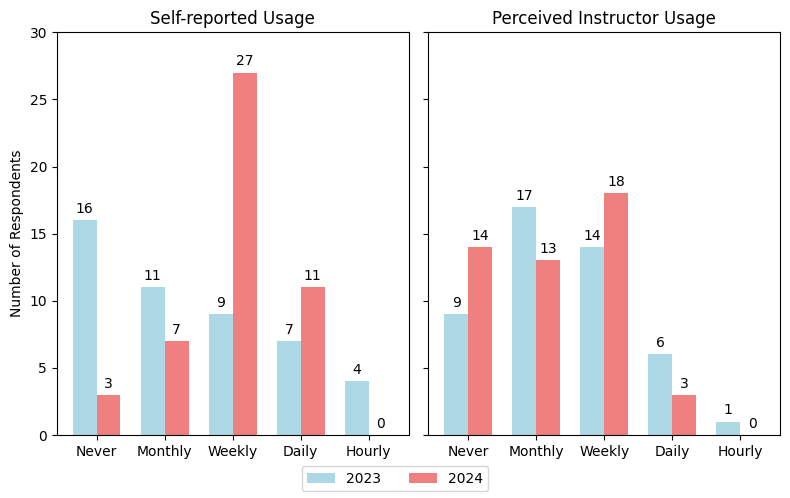}
    \caption{Participant self-reported frequency of usage of GenAI tools compared to their perceived instructor usage}
    \label{fig:enter-label}
\end{figure}

Students ranked common computing help-seeking resources in order of preference. Table ~\ref{tab:frequent-percent} presents the percentage point differences in the frequency of usage reported by computing students across seven resources between 2023 and 2024. The survey results for 2024 indicate that GenAI tools, specified in the survey as ChatGPT or similar GenAI tools, experienced the most substantial change in frequency of usage. 

In 2023, students ranked GenAI fourth for most relied-upon resource, behind online search, peers, and class forums. In 2024, GenAI ranked second, with first being online search. Online search maintained a marginal lead, with just 2.08\% of students relying more frequently on the internet than on GenAI. In 2024, 93.75\% participants reported using ChatGPT or similar GenAI tools at least once a month, compared to 2023's 65.96\% of participants. The percentage of students relying on ChatGPT to meet their help-seeking needs on a weekly basis has increased substantially (+37.10\%), paralleled by a decrease in students who never use the resource (-27.79\%). Only 6.25\% of students reported they do not use ChatGPT, a decrease from the previous year's 34.04\%. On the other hand, hourly usage of ChatGPT decreased by 8.51\%, which could be due to a variety of reasons, such as survey sample size and students' tendency to underestimate their usage~\cite{prather2024widening}. Table~\ref{tab:frequent-percent} also shows a decline in student usage of human help-seeking resources, with daily help-seeking from friends decreasing by 8.91\%; students shifted to help-seeking from friends more often on a weekly basis (+20.12\%). Instructor and TA usage also declined slightly in terms of daily reliance, with a 10.02\% increase of students reporting they never sought help from instructors. GenAI resources like GitHub Copilot remained the least frequently-relied upon, despite a 5.93\% increase in overall use.

\section{Discussion}

As generative AI usage among students has evolved, early findings suggested there was a divide between students who used ChatGPT frequently while others rarely used these tools~\cite{smith2024early, hou2024effects, prather2024widening}. However, the results presented in this paper suggest that the initial divide between early adopters and late adopters is shrinking. A rapidly growing majority of computing students have adopted ChatGPT as their primary choice of help, and the resource usage of GenAI tools are now nearly on par with the previously dominant internet search. Given that GenAI has risen to the rank of second most relied-upon resource in the span of a year, it may be set to surpass online search in coming years. This shift over time suggests a fundamental change in how computing students are seeking help. Students are increasingly prioritizing the instantaneous and personalized help of GenAI tools over traditional search-based information retrieval, influencing the ways in which they learn and synthesize new information without human intervention. 

The decline of hourly and daily reliance on GenAI also indicates other potential phenomena at play. This decline could be due to the novelty effect or increased restrictions on token limits for ChatGPT~\cite{openAI_2024}, which prevents users from frequently querying ChatGPT for free. This decline could also be a sign of students becoming desensitized to the extent of their GenAI use, resulting in underestimations of their reliance~\cite{prather2024widening}. 

Student reliance on peers and instructors has also slightly declined. Engagement with peers and instructors has become more sporadic, the effects of which could affect peer-to-peer relationships, classroom interactions, and the design of course instruction. As GenAI grows in help-seeking value, students may find that the social and temporal costs of asking for help from their peers and instructors are no longer worth it. Past research has observed that students before GenAI highly preferred peer-to-peer help~\cite{doebling2021patterns}. Future research is urgently needed to explore this changing trade-off and its potential impact.

While the gap between students who use GenAI and those who do not is shrinking, more research in this area is needed to examine whether students are using these tools in the same manner or deriving the same benefits. Recent findings suggest that this may \textit{NOT} be the case~\cite{prather2024widening} but more research is needed. In other words, although the usage gap is closing,  performance gaps may remain.

\section{Limitations} 
The limitations of this study include the generalizability of results due to smaller sample sizes and focus on North American participants. Despite these limitations, our findings align with similar research on students' use of ChatGPT~\cite{smith2024early}. Additionally, surveys were cross-sectional and not conducted within the same group of participants. Future research could assess changes over time within the same group to better understand the longitudinal differences in resource reliance. It is also possible that students under-reported their GenAI usage, as suggested by Prather et al.~\cite{prather2024widening}.

\bibliographystyle{ACM-Reference-Format}
\bibliography{sample-base}


\begin{thebibliography}{7}


\ifx \showCODEN    \undefined \def \showCODEN     #1{\unskip}     \fi
\ifx \showDOI      \undefined \def \showDOI       #1{#1}\fi
\ifx \showISBNx    \undefined \def \showISBNx     #1{\unskip}     \fi
\ifx \showISBNxiii \undefined \def \showISBNxiii  #1{\unskip}     \fi
\ifx \showISSN     \undefined \def \showISSN      #1{\unskip}     \fi
\ifx \showLCCN     \undefined \def \showLCCN      #1{\unskip}     \fi
\ifx \shownote     \undefined \def \shownote      #1{#1}          \fi
\ifx \showarticletitle \undefined \def \showarticletitle #1{#1}   \fi
\ifx \showURL      \undefined \def \showURL       {\relax}        \fi
\providecommand\bibfield[2]{#2}
\providecommand\bibinfo[2]{#2}
\providecommand\natexlab[1]{#1}
\providecommand\showeprint[2][]{arXiv:#2}

\bibitem[Doebling and Kazerouni(2021)]%
        {doebling2021patterns}
\bibfield{author}{\bibinfo{person}{Augie Doebling} {and} \bibinfo{person}{Ayaan~M. Kazerouni}.} \bibinfo{year}{2021}\natexlab{}.
\newblock \showarticletitle{Patterns of Academic Help-Seeking in Undergraduate Computing Students}. In \bibinfo{booktitle}{\emph{Proceedings of the 21st Koli Calling International Conference on Computing Education Research}}.
\newblock
\showISBNx{9781450384889}


\bibitem[Hou et~al\mbox{.}(2024)]%
        {hou2024effects}
\bibfield{author}{\bibinfo{person}{Irene Hou}, \bibinfo{person}{Sophia Mettille}, \bibinfo{person}{Owen Man}, \bibinfo{person}{Zhuo Li}, \bibinfo{person}{Cynthia Zastudil}, {and} \bibinfo{person}{Stephen MacNeil}.} \bibinfo{year}{2024}\natexlab{}.
\newblock \showarticletitle{The Effects of Generative AI on Computing Students’ Help-Seeking Preferences}. In \bibinfo{booktitle}{\emph{Proceedings of the 26th Australasian Computing Education Conference}} \emph{(\bibinfo{series}{ACE 2024})}.
\newblock
\urldef\tempurl%
\url{https://doi.org/10.1145/3636243.3636248}
\showDOI{\tempurl}


\bibitem[OpenAI(2024)]%
        {openAI_2024}
\bibfield{author}{\bibinfo{person}{OpenAI}.} \bibinfo{year}{2024}\natexlab{}.
\newblock
\newblock
\urldef\tempurl%
\url{https://platform.openai.com/docs/guides/rate-limits/usage-tiers}
\showURL{%
\tempurl}


\bibitem[Padiyath et~al\mbox{.}(2024)]%
        {padiyath2024insights}
\bibfield{author}{\bibinfo{person}{Aadarsh Padiyath}, \bibinfo{person}{Xinying Hou}, \bibinfo{person}{Amy Pang}, \bibinfo{person}{Diego Viramontes~Vargas}, \bibinfo{person}{Xingjian Gu}, \bibinfo{person}{Tamara Nelson-Fromm}, \bibinfo{person}{Zihan Wu}, \bibinfo{person}{Mark Guzdial}, {and} \bibinfo{person}{Barbara Ericson}.} \bibinfo{year}{2024}\natexlab{}.
\newblock \showarticletitle{Insights from Social Shaping Theory: The Appropriation of Large Language Models in an Undergraduate Programming Course}. In \bibinfo{booktitle}{\emph{Proceedings of the 2024 ACM Conference on International Computing Education Research-Volume 1}}. \bibinfo{pages}{114--130}.
\newblock


\bibitem[Prather et~al\mbox{.}(2023)]%
        {prather2023robots}
\bibfield{author}{\bibinfo{person}{James Prather}, \bibinfo{person}{Paul Denny}, \bibinfo{person}{Juho Leinonen}, \bibinfo{person}{Brett~A Becker}, \bibinfo{person}{Ibrahim Albluwi}, \bibinfo{person}{Michelle Craig}, \bibinfo{person}{Hieke Keuning}, \bibinfo{person}{Natalie Kiesler}, \bibinfo{person}{Tobias Kohn}, \bibinfo{person}{Andrew Luxton-Reilly}, {et~al\mbox{.}}} \bibinfo{year}{2023}\natexlab{}.
\newblock \showarticletitle{The robots are here: Navigating the generative ai revolution in computing education}.
\newblock In \bibinfo{booktitle}{\emph{Proceedings of the 2023 Working Group Reports on Innovation and Technology in Computer Science Education}}. \bibinfo{pages}{108--159}.
\newblock


\bibitem[Prather et~al\mbox{.}(2024)]%
        {prather2024widening}
\bibfield{author}{\bibinfo{person}{James Prather}, \bibinfo{person}{Brent~N Reeves}, \bibinfo{person}{Juho Leinonen}, \bibinfo{person}{Stephen MacNeil}, \bibinfo{person}{Arisoa~S Randrianasolo}, \bibinfo{person}{Brett~A. Becker}, \bibinfo{person}{Bailey Kimmel}, \bibinfo{person}{Jared Wright}, {and} \bibinfo{person}{Ben Briggs}.} \bibinfo{year}{2024}\natexlab{}.
\newblock \showarticletitle{The Widening Gap: The Benefits and Harms of Generative AI for Novice Programmers}. In \bibinfo{booktitle}{\emph{Proceedings of the 2024 ACM Conference on International Computing Education Research - Volume 1}} \emph{(\bibinfo{series}{ICER '24})}.
\newblock
\showISBNx{9798400704758}
\urldef\tempurl%
\url{https://doi.org/10.1145/3632620.3671116}
\showDOI{\tempurl}


\bibitem[Smith et~al\mbox{.}(2024)]%
        {smith2024early}
\bibfield{author}{\bibinfo{person}{C.~Estelle Smith}, \bibinfo{person}{Kylee Shiekh}, \bibinfo{person}{Hayden Cooreman}, \bibinfo{person}{Sharfi Rahman}, \bibinfo{person}{Yifei Zhu}, \bibinfo{person}{Md~Kamrul Siam}, \bibinfo{person}{Michael Ivanitskiy}, \bibinfo{person}{Ahmed~M. Ahmed}, \bibinfo{person}{Michael Hallinan}, \bibinfo{person}{Alexander Grisak}, {and} \bibinfo{person}{Gabe Fierro}.} \bibinfo{year}{2024}\natexlab{}.
\newblock \showarticletitle{Early Adoption of Generative Artificial Intelligence in Computing Education: Emergent Student Use Cases and Perspectives in 2023}. In \bibinfo{booktitle}{\emph{Proc. of Innovation and Technology in Computer Science Education V. 1}}.
\newblock
\showISBNx{9798400706004}


\end{thebibliography}


\end{document}